\documentclass[superscriptaddress,aps,pra,twocolumn,showpacs,nofootinbib,longbibliography,notitlepage,floatfix]{revtex4-2}
\usepackage{amsmath,amssymb,amsthm}
\usepackage[colorlinks=true,citecolor=blue,urlcolor=blue]{hyperref}
\usepackage[pdftex]{graphicx}
\usepackage{times,txfonts}
\usepackage{braket}
\usepackage{color}
\usepackage{natbib}
\usepackage{amsmath,blkarray}
\usepackage{mathtools}
\usepackage{ulem}
\usepackage{latexsym}
\usepackage{tabularx, booktabs}
\usepackage{graphics,epstopdf}
\usepackage{graphicx}
\usepackage{float}
\usepackage{amsfonts}

\newcommand{\be}{\begin{equation}}
\newcommand{\ee}{\end{equation}}
\newcommand{\ba}{\begin{eqnarray}}
\newcommand{\ea}{\end{eqnarray}}

\usepackage{multirow}
\usepackage{appendix}
\usepackage{url}

\begin{document}
\title{Shortcuts to Adiabatic Soliton Compression in Active Nonlinear Kerr Media}
%
%

\author{Yingjia Li}
\affiliation{Department of Physics, Shanghai University, 200444 Shanghai, China}
\affiliation{Department of Physical Chemistry, University of the Basque Country UPV/EHU, Apartado 644, 48080 Bilbao, Spain}

\author{Koushik Paul}
\email{koushikpal09@gmail.com}
\affiliation{Department of Physical Chemistry, University of the Basque Country UPV/EHU, Apartado 644, 48080 Bilbao, Spain}
\affiliation{EHU Quantum Center, University of the Basque Country UPV/EHU, Barrio Sarriena, s/n, 48940 Leioa, Spain}

\author{David Novoa}
\affiliation{Department of Communications Engineering, Engineering School of Bilbao, University of the Basque Country (UPV/EHU), 48013 Bilbao, Spain}
\affiliation{EHU Quantum Center, University of the Basque Country UPV/EHU, Barrio Sarriena, s/n, 48940 Leioa, Spain}
\affiliation{IKERBASQUE, Basque Foundation for Science, 48009 Bilbao, Spain}

\author{Xi Chen}
\email{xi.chen@ehu.eus}
\affiliation{Department of Physical Chemistry, University of the Basque Country UPV/EHU, Apartado 644, 48080 Bilbao, Spain}
\affiliation{EHU Quantum Center, University of the Basque Country UPV/EHU, Barrio Sarriena, s/n, 48940 Leioa, Spain}


\begin{abstract}
%
We implement variational shortcuts to adiabaticity for optical pulse compression in an active nonlinear Kerr medium with distributed amplification and spatially varying dispersion and nonlinearity. Starting with the hyperbolic secant ansatz, we employ a variational approximation to systematically derive dynamical equations, establishing analytical relationships linking the amplitude, width, and chirp of the pulse. Through the inverse engineering approach, we manipulate the distributed gain/loss, nonlinearity and dispersion profiles to efficiently compress the optical pulse over a reduced distance with high fidelity. In addition, we explore the dynamical stability of the system to illustrate the advantage of our protocol over conventional adiabatic approaches. Finally, we analyze the impact of tailored higher-order dispersion on soliton self-compression and derive physical constraints on the final soliton width for the complementary case of soliton expansion. The broader implications of our findings extend beyond optical systems, encompassing areas such as cold-atom and magnetic systems highlighting the versatility and relevance of our approach in various physical contexts.

\end{abstract}
\maketitle
\section{Introduction}

Optical solitons are stationary solutions of the nonlinear Schr\"{o}dinger equation arising from the perfect balance between dispersive and nonlinear properties of the medium. They have been extensively studied due to their fundamental properties and practical applications in many disparate fields such as optical communications, nonlinear frequency conversion, laser science, quantum information processing or ultrafast spectroscopy \cite{Agrawal,solitonrev,sch}. Among the different soliton-related effects of relevance, pulse compression in nonlinear media such as optical fibers stands out since it enables access to ultrashort pulse duration and high peak power by carefully controlling the properties of the fiber. To do this, there are several widely developed techniques including fiber-grating compression \cite{shankapl,ForkOL,TomlinsonJOSA}, soliton-effect compression \cite{Gordonprl,Tomlinsonol,Ahmedieee,FosterOE}, and adiabatic pulse compression \cite{Kuehl,Smith1989,ChernikovJOSA,Chernikovol,ChernkovEL,Mamyshevprl}.
While both soliton-effect and adiabatic pulse compression involve the careful manipulation of the in-fiber propagation dynamics of optical pulses, they use different approaches. In detail, soliton-effect self-compression is achieved by balancing the dispersion and nonlinearity of the fiber, allowing the pulse to decrease its duration while maintaining its shape \cite{Pelusi}. On the other hand, adiabatic pulse compression utilizes chirped pulses that are gradually compressed through dispersion-compensation elements, resulting in either shorter and higher intensity outcoupled pulses \cite{CPA55}, or minimizing the associated pedestals with specially designed dispersion profiles \cite{DDFad}. 

Methodologically, adiabatic soliton compression involves the variation of fiber parameters such as dispersion and nonlinearity, and/or the soliton energy through gain and/or loss, slowly relative to the characteristic soliton length \cite{Chernikovol}. While adiabatic soliton compression has many advantages, such as preserving the shape and quality of the pulse, there are also some potential shortcomings to consider \cite{Fatemi,Travers,Lagsgaard,Turitsynprl,Turitsyn}. For instance, it requires relatively long fiber lengths to achieve significant compression ratios. In addition, nonlinear effects such as four-wave mixing and stimulated Raman scattering can also introduce distortions in the pulse shape, as well as unwanted noise, thereby reducing the achievable compression ratio and overall coherence. Therefore, strategies for nonadiabatic compression of optical pulses are desirable in practical applications where tight spatial constraints for miniaturization and integration are enforced \cite{josb1994,Koushik,Kong,fiber_non}.

Here we focus on shortcuts to adiabatic compression of optical pulses in an active nonlinear Kerr medium by judiciously designing several parameters of the system such as distributed amplification, varying dispersion, or controllable nonlinearity. The methods of shortcuts to adiabaticity (STA) \cite{STArev} manage to accelerate various adiabatic processes in atomic, molecular and optical physics among many other fields. In order to deal with the nonlinear case, where a dynamical invariant does not exist, we have recently proposed the use of inverse engineering along with the variational approximation for designing the STA control of ensembles of interacting atoms or matter-wave solitons \cite{Zoller,Li2016,Huangchaos,Huangpra}. In an analogous fashion, the variational approximation \cite{Anderson,Borisbook}  and self-similarity \cite{Kruglovprl} thus provide the dynamical equation for capturing the nonlinear propagation dynamics of optical pulses in terms of amplitude, width and the chirp. Furthermore, its combination with inverse engineering allows to design high-fidelity compression of optical pulses beyond the adiabatic criteria provided suitable boundary conditions are fulfilled. Although this work covers the application of our method to pulse compression in optical fiber systems, it might be straightforwardly extended to other platforms featuring e.g. cubic-quintic (high-order) \cite{ZhangPRA} or nonlocal nonlinearities \cite{Olebang,Borispra,Agarwalpre,Sabari}.

The paper is organized as follows. In Sec. \ref{model and formula} we present both the physical model and STA method. With this, we introduce in Sec. \ref{sta} the inverse engineering of controllable parameters in active optical fibers for efficient pulse compression beyond the adiabatic condition. In Sec. \ref{discussion}, we discuss the efficiency and stability of our methods, showing the superior performance of STA compared to the conventional adiabatic one. Finally, the main conclusions of this study are briefly summarized in Sec. \ref{conclusion}. 

\section{Physical Model and STA Method}
 
\label{model and formula}

The propagation of an optical pulse through a Kerr nonlinear medium such as an optical fiber-based system with distributed gain and varying dispersion and nonlinearity, is described by the modified nonlinear Schr\"{o}dinger equation (NLSE) \cite{sch} as follows:
\begin{equation}
	i\frac{\partial u}{\partial z}+\frac{1}{2}\beta_2(z)\frac{\partial^{2} u}{\partial t^{2}}+\gamma(z)|u|^{2}u=ig(z)u,
	\label{sch}
\end{equation}
where we have introduced dimensionless variables using physical units (denoted by tildes):
\begin{equation}\nonumber
	u=\tilde{u}/\sqrt{P_0},\quad z=\tilde{z}/L_D, \quad t=(\tilde{t}-\tilde{z}/v_g)/T_0.
	\end{equation}
Here, $\tilde{u}$, $\tilde{z}$, and $\tilde{t}$ represent the complex electric field envelope of the optical pulse, the longitudinal coordinate in the fiber, and time, respectively. The parameters include $P_0$, the peak power of the pulse, and $v_g$, the group velocity at the central wavelength. The dispersion length is denoted by $L_D=T_0^2/|\beta_2|$, where $T_0$ is the initial pulse width, and $\beta_2(z)$ represents the spatially-varying group-velocity dispersion (GVD). The nonlinear fiber parameter, $\gamma(z)=\tilde{\gamma}(z)P_0L_D$, characterizes the Kerr nonlinearity \cite{sch}, where the change in refractive index is proportional to $|u|^2$. Note that this model's implementation spans various techniques, including tapered photonic crystal fibers \cite{Tapers}. On the right side of the equation, the term $ig(z)u$ accounts for stochastic gain distributed across the entire active fiber length. This formulation provides a comprehensive framework for studying pulse dynamics in diverse complex optical systems.

For convenience in qualitative analysis and numerical simulations, we integrate the gain term into the nonlinear coefficient using the following transformation: $u(z,t) = A(z,t)$ $\exp\left[\int^{z}_{0}g(z')dz'\right]$. In this framework, Eq. (\ref{sch}) transforms into:
\begin{equation}
	i\frac{\partial A}{\partial z}+\frac{1}{2}\beta_2(z)\frac{\partial^{2}A}{\partial t^{2}}+G(z)\gamma(z)|A|^{2}A=0,
	\label{sch_new}
\end{equation}
where $G(z)=\exp\left[2\int^{z}_{0}g(z')dz'\right]$.  
This modification accounts for the amplification of the nonlinear coefficient due to the presence of gain within the fiber.

For the analytical treatment hereafter, we adopt the following ansatz, describing the bright solitary wave solutions of a passive NLSE:
\begin{equation}
	A(z,t) = A_0(z)\mathrm{sech} \left[\frac{t}{a(z)}\right]\exp[ib(z)t^{2}+i\phi(z)],
	\label{ansatz}
\end{equation}
where $A_0(z), a(z), b(z)$ and $\phi(z)$ represent the amplitude, width, chirp and phase, respectively. These are real functions, 
and $A_0(z)=\sqrt{1/2a(z)}$ is determined by ensuring that the electric field is properly normalized. 
By leveraging the Lagrangian formalism (refer to Appendix~\ref{va}), 
an Ermakov-like equation is derived, governing the variations in the soliton parameters:
\begin{equation}
	\frac{d}{dz}\left[\frac{1}{\beta_2(z)}\frac{da}{dz}\right]=\frac{4\beta_2(z)}{\pi^{2}a^{3}}-\frac{2\gamma(z)G(z)}{\pi^{2}a^{2}}\equiv-\frac{\partial V}{\partial a}.
	\label{ermakov}
\end{equation}
Henceforth, we omit the $z$-dependence for brevity, e.g. $a(z) \equiv a$, unless specified.
It's noteworthy that Eq. (\ref{ermakov}) is analogous to Newton’s equation of motion, delineating the dynamics of a particle with variable mass $1/\beta_2(z)$ in an effective potential $V$ \cite{variablemass}. Consequently, we can utilize Eq. (\ref{ermakov}) to inversely engineer the soliton parameters for designing STA with appropriate boundary conditions.
 
\section{Inverse Engineering}
\label{sta}
Optical pulse compression involves diverse approaches centered around the manipulation of distinct parameters. These techniques encompass tailoring both the GVD and nonlinearity for compression \cite{nonlinear2003}, as well as controlling the distributed gain along the fiber \cite{josb1996}. In what follows, we will analyze the impact of engineering each of these properties, adhering to the principles outlined by STA-based strategies.


\subsection{Distributed gain}
\label{gain}

Distributed gain in optical fibers is typically achieved by introducing dopant elements, such as erbium or ytterbium, into the core. The spatial distribution of rare-earth dopants can be strategically designed to counteract the effects of dispersion-induced pulse broadening during propagation, ultimately enabling efficient soliton compression.

Building upon this inspiration, our objective is to develop STA protocols for soliton compression by devising an appropriate 
distributed gain with constant GVD and Kerr nonlinearity. In this case, Eq. (\ref{ermakov}) is simplified as
\begin{equation}
\label{reduced-Ermakov}
 \ddot{a} = \frac{4 \beta^2_2}{\pi^2 a^3} - \frac{2 \beta_2 \gamma G(z)}{\pi^2 a^2}.
\end{equation}
Further derivatives with respect to $z$ lead to:
\begin{equation}
\dddot{a} = - \frac{12{\beta^2_2}\dot a}{\pi^2 a^4}+\frac{4{\beta_2}\gamma G\dot a}{\pi^2 a^3}-\frac{4{\beta_2}\gamma G g}{\pi^2 a^2}.
\label{tod_derivate}
\end{equation}
Based on Eq. (\ref{reduced-Ermakov}), this system can be likened to the perturbative Kepler problem \cite{Li2016}, where adiabaticity implies a fictitious particle of variable mass remaining at the minimum of the effective potential with zero velocity,  
e.g., $ \ddot{a}=-\partial V/ \partial a=0$ and $\dot{a}=0$. In this scenario, Eq. (\ref{reduced-Ermakov}) yields the the following adiabatic reference:
\begin{equation}
a_c(z) = \frac{2 \beta_2}{\gamma G(z)}.
 \label{ad}
\end{equation}
This reference determines $a(0)= a_c (0)$ and  $a(z_f)= a_c (z_f)$, especially when the adiabatic protocol $G(z)= e^{2 g_0 z}$ with constant $g_0$ is chosen for comparison. Meanwhile,
the boundary conditions $\dot{a}(0)= \dot{a}(z_f) =0 $ and $\ddot{a}(0)= \ddot{a} (z_f) = 0$ should be embraced. Consequently, all these boundary conditions guarantee a swift transition from the initial width $a(0)$ to the desired final width $a(z_f)$, 
maintaining stationary initial and final states. Moreover, taking into account the transformation, $G(z)=\exp\left[2\int^{z}_{0}g(z')dz'\right]$ used earlier, 
the designed $g(z)$ has to be consistent with adiabatic reference at the edges. Therefore, the boundary conditions, 
$\dddot{a} (0)= \dddot{a}_c (0)$ and $\dddot{a} (0)= \dddot{a}_c (z_f)$ are further established, 
by solving
\begin{equation}
\label{dddotcondition}
\dddot{a}_c = -\frac{4{\beta_2}\gamma G g_0}{\pi^2 a^2},
\end{equation}
as derived from Eq. (\ref{tod_derivate}) with the adiabatic protocol within the adiabatic condition. 

\begin{figure}[t]
\centering
\hspace*{-.5cm}\begin{minipage}{0.45\linewidth}
  \includegraphics[width=\linewidth]{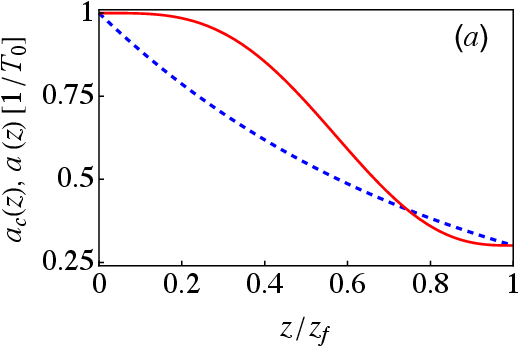}
\end{minipage}%
\hspace*{.5cm}\begin{minipage}{0.45\linewidth}
  \centering
  \includegraphics[width=\linewidth]{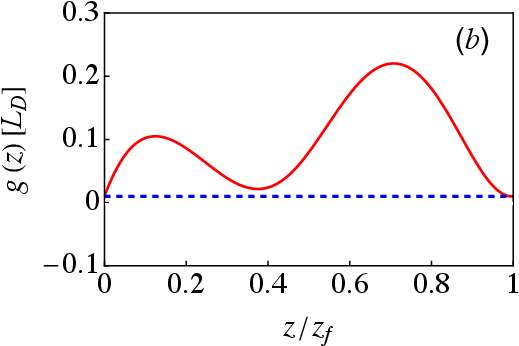}
\end{minipage}
\hspace*{-0.15cm}\begin{minipage}{0.48\linewidth}
  \centering
  \includegraphics[width=\linewidth]{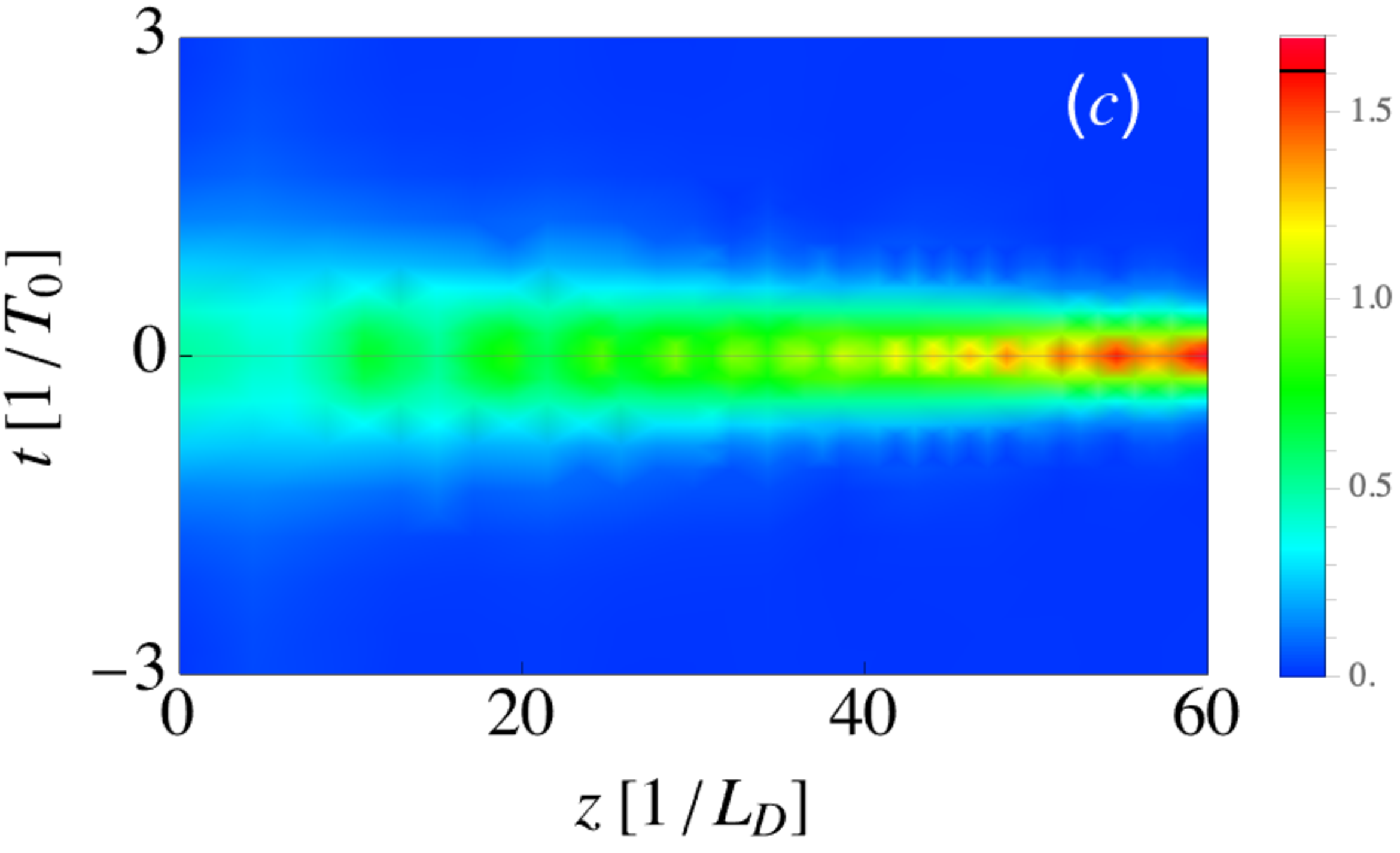}
\end{minipage}%
\hspace*{.2cm}\begin{minipage}{0.47\linewidth}
  \centering
  \includegraphics[width=\linewidth]{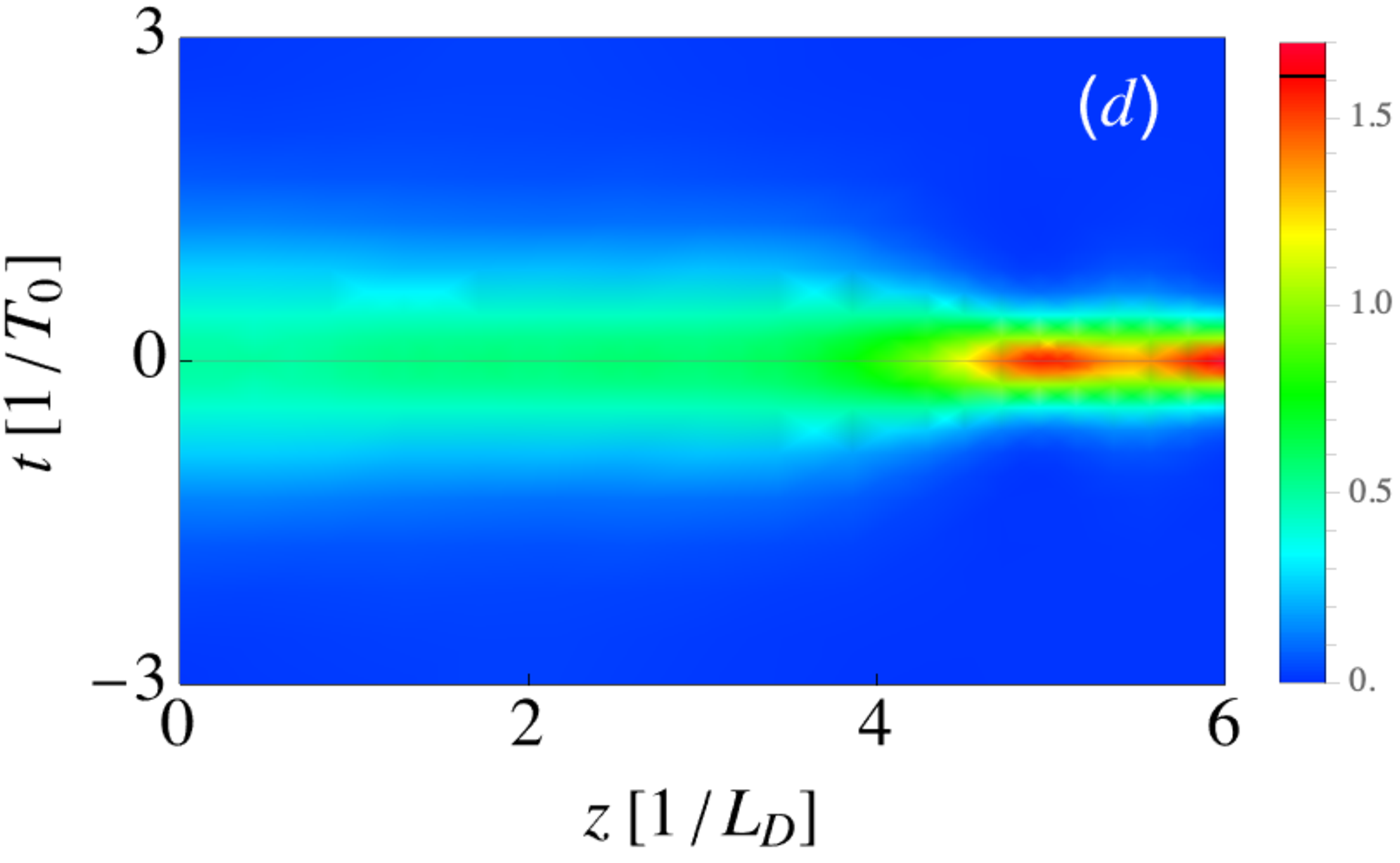}
\end{minipage}
\caption{(a) A comparison between $a(z)$ (red-solid line) obtained using inverse engineering with a final fiber length of $z_f=6$ and the adiabatic reference (blue-dashed line) with $z_f=60$. (b) The corresponding distributed gain $g(z)$ designed via inverse engineering (red-solid line) compared with the adiabatic protocol with constant $g_0$ (blue-dashed line). (c) and (d) illustrate the spatio-temporal evolution of soliton wave packets, respectively, in both the adiabatic and STA protocols. Other parameters are: $g_0=0.01$, $\beta_2=1$ and $\gamma=2$.}
\label{evolutionone}
\end{figure}

In the context of fixed boundary conditions, we can choose a straightforward polynomial ansatz given by 
\begin{equation}
a(z) = \sum_{j=0}^{n}c_{j}z^{j},
\label{polynomial ansatz}
\end{equation}
  to interpolate the function of $a(z)$ for $n=7$. The coefficients $c_j$ can be determined by combining the ansatz with the provided boundary conditions. Subsequently, with the interpolated function of $a(z)$, we can derive the distributed gain $g(z)$ from Eq. (\ref{reduced-Ermakov}). As depicted in Fig.~\ref{evolutionone}(a), the soliton width is compressed from $a_c(0)$ to $a_c(z_f)$ in both adiabatic and STA cases. However, the evolution of $a(z)$ designed through STA (solid-red line) deviates from the adiabatic reference $a_{c}(z)$ (blue-dashed line). Remarkably, the propagation distance is reduced by a factor of 10 in the STA protocol ($z_f=6$), while achieving a similar pulse width reduction as the adiabatic protocol requires an extensive $z_f=60$ with the small constant $g_0$. 

Figure \ref{evolutionone}(b) also shows the distributed gain $g(z)$, corresponding to the design from STA and adiabatic protocols. From a mathematical standpoint, the propagation distance $z_f$ can be chosen to be extremely short in soliton self-compression. However, this approach comes with certain challenges. As the fiber length is decreased below a certain point (not depicted here), the designed $g(z)$ may take on negative values. This signifies a complex gain and loss profile as well as implying the cost for such intricate design of the gain profile using STA in short fibers.  Moreover, in the case of very short fiber lengths, there exists the possibility of the distributed gain rapidly reaching saturation which might hinder the achievement of further effective compression. 
On the contrary, Eq. (\ref{dddotcondition}) clearly indicates that the STA protocol has limitations when it comes to long fiber lengths. 
From  Eqs. (\ref{reduced-Ermakov}) and (\ref{tod_derivate}), it is evident that $\dddot a_c(z_f)$ increase with increasing $z_f$ which introduces time-dependency in the forces within Newton's equations of motion. Therefore when we impose $\dddot{a}(0)=\dddot{a}_c (0)$ and $\dddot{a}(z_f)=\dddot{a}_c (z_f)$ the effective potential becomes dissipative in nature. As a consequence, the designed STA protocol becomes invalid for controlling soliton compression when $|\dddot{a}_c| >1$. This particular property makes it different from the soliton compression by using inverse engineering with controllable nonlinearity. In short, the application of the STA protocol can achieve robust and rapid soliton compression only when the fiber length is moderately short, following certain conditions, such as $z_f=10$, $|\dddot a(z_f)|<0.015$ to maintain the conservative nature of the evolution.

In Fig.~\ref{evolutionone} (c), the spatio-temporal propagation of electric field $|A(z, t)|^2$ is depicted under both adiabatic (panel (c)) and STA protocols (panel (d)). As previously discussed, the self-compression dynamics occur significantly faster in the system designed with the STA protocol, allowing for the similar performance with moderate fiber lengths, specifically $z_{f} = 6$. As a reference, we have chosen the adiabatic protocol, represented by $G(z)= e^{2 g_0 z}$ with a constant $g_0=0.01$, ensuring that the adiabatic condition is met. 
Furthermore, we shall emphasize that the values of $g(0)$ and $g(z_f)$ tend to the value of $g_0$
corresponding to the adiabatic reference when the boundary conditions, including the third order derivative at the edge, are carefully set.


\subsection{Group-velocity dispersion}
\label{GVD}

\begin{figure}[t]
\centering
\hspace*{-.5cm}\begin{minipage}{0.45\linewidth}
  \includegraphics[width=\linewidth]{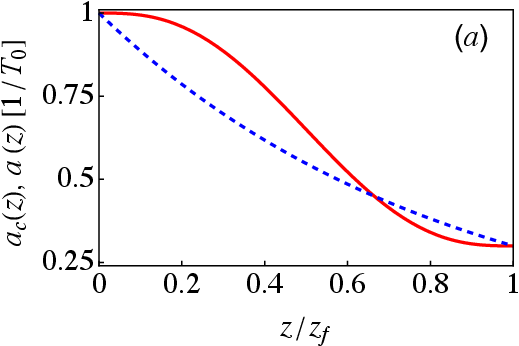}
\end{minipage}%
\hspace*{.5cm}\begin{minipage}{0.45\linewidth}
  \centering
  \includegraphics[width=\linewidth]{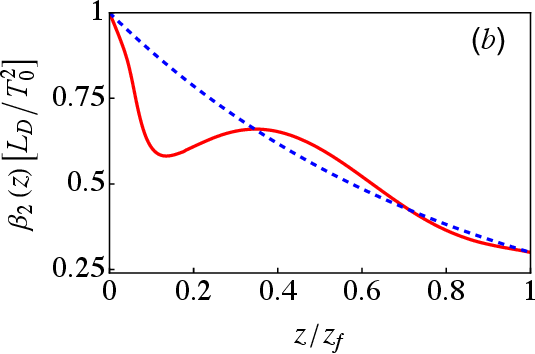}
\end{minipage}
\hspace*{-0.15cm}\begin{minipage}{0.48\linewidth}
  \centering
  \includegraphics[width=\linewidth]{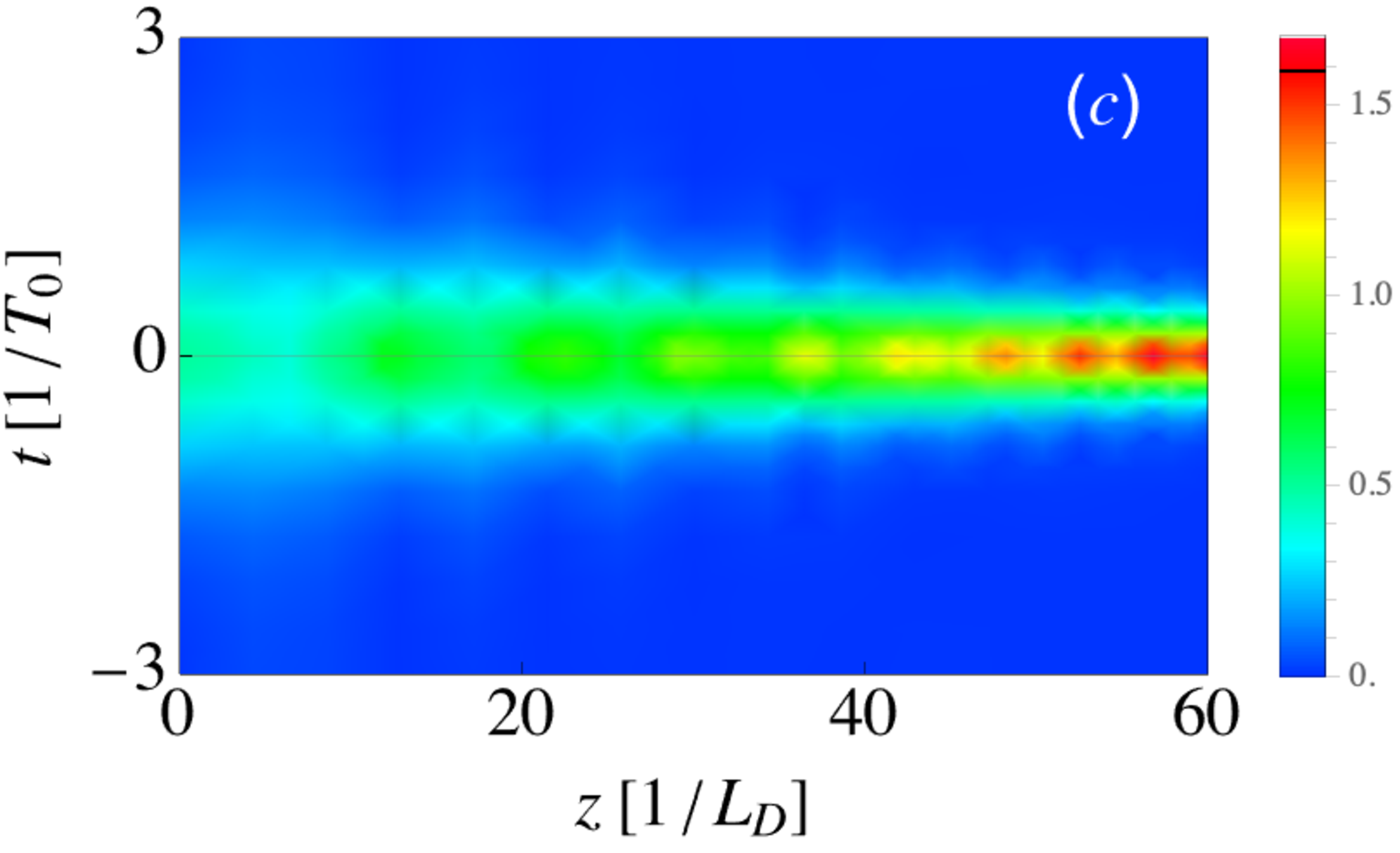}
\end{minipage}%
\hspace*{.2cm}\begin{minipage}{0.47\linewidth}
  \centering
  \includegraphics[width=\linewidth]{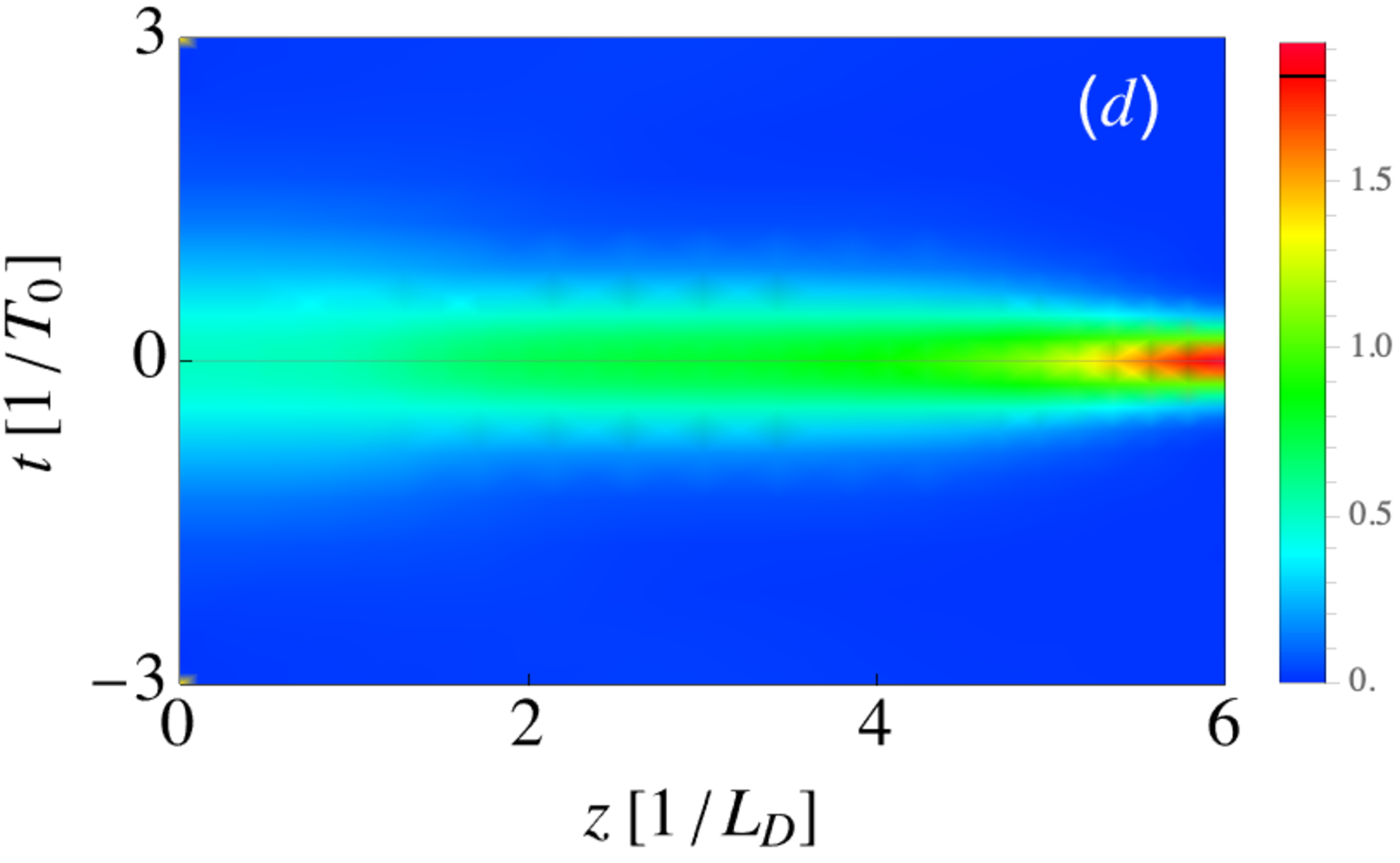}
\end{minipage}
\caption{(a) A comparison between $a(z)$ (red-solid line) obtained using inverse engineering with a final fiber length of $z_{f}=6$ and the adiabatic reference (blue-dashed line) with $z_{f}=60$. (b) 
The corresponding GVD $\beta_2(z)$ is designed through inverse engineering with the STA protocol shown in red-solid line, while the blue-dashed line represents the GVD profile with adiabatic protocol. (c) and (d) 
illustrate the spatio-temporal evolution of soliton wave packets in the adiabatic and STA protocols. Here the constant Kerr nonlinearity is set to $\gamma=2$, $\beta_2 (0)=1$ and $\beta_2 (z_f)=0.3$.}
\label{reac}
\end{figure}

Dispersion management serves as a potent tool for controlling and shaping the dynamics of optical pulses as they propagate through optical fibers. 
A particularly useful implementation of this strategy is the dispersion-decreasing fiber (DDF) \cite{DDLfiber}. 
These specialized fibers can be fabricated through tapering or by direct modifying the refractive index profile. This manipulation involves engineering the core and cladding materials, as well as doping specific elements to achieve the desired effects on dispersion. 
However, if the fiber length is too long, nonlinear effects such as excessive self-phase modulation and stimulated Raman scattering can accumulate and distort the compressed pulses. To overcome this limitation, we aim to accelerate the adiabatic process and reduce the fiber length by implementing the concept of inverse engineering to construct the STA protocol.

To simplify, let's focus on the passive scenario with no gain/loss, a constant Kerr nonlinearity, and variable GVD, denoted as $\beta_2(z)$. In this case,
Eq. (\ref{ermakov}) can be simplified to:
\begin{equation}
\label{reduced-Ermakov-2}
 \ddot{a} = \frac{\dot{\beta}_2}{\beta_2}\dot a + \frac{4 \beta^2_2} {\pi^2 a^3} - \frac{2 \beta_2 \gamma}{\pi^2 a^2}.   
\end{equation}
Notably, in this situation,  we have no gain or loss, meaning $g=0$, and thus $G=1$. 
We can define the initial and final soliton widths by establishing the following boundary conditions: $a(0)=a_c (0)$, $a(z_f)=a_c(z_f)$, $\dot{a}(0) = \dot{a}(z_{f})=0$, $\ddot{a}(0) = \ddot{a}(z_{f})=0$. These conditions ensure that the initial and final soliton widths remain unchanged and distortion-free.
Similar to Eq. (\ref{ad}), the values of $a_c(0)$ and $a_c(z_f)$ can be calculated from the adiabatic reference,
\begin{equation}
a_c (z)= 2 \beta_2 (z)/ \gamma,
\end{equation}
where $\beta_2 (z)$ is given by
\begin{equation}
\beta_2(z)=\exp\left(-\frac{\ln\beta}{z_f}z\right).
\label{acbeta}
\end{equation}
This particular choice of $\beta_2(z)$ is found to offer rapid-adiabatic compression, fulfilling the adiabatic condition \cite{DDFad}:
\begin{equation}
\label{adiabaticcondition-beta}
 \left|\frac{1}{\beta_2(0)}\frac{d\beta_2 (z)}{dz} \right|\ll \frac{1}{z_f}.
\end{equation}
In this context, soliton propagating in a fiber with exponentially decreasing dispersion can achieve adiabatic compression due to the monotonic decrease of $\beta_2(z)$ from the initial value $\beta_2(0)$ to the 
final one $\beta_2(z_f)$. This decrease is determined by the dispersion ratio $\beta=\beta_2(0)/\beta_2(z_f)$. Among various dispersion-decreasing profiles, including 
linear, hyperbolic, Gaussian, and exponential, it has been observed that the latter offers the fastest adiabatic compression, and DDFs with exponential profiles can be practically implementable \cite{DDFexp}.

Once the boundary conditions are defined with the help of adiabatic reference, the trajectory of $a(z)$ can be represented as Eq. (\ref{polynomial ansatz}) for $n=5$, as used before. As a result, Fig. \ref{reac}(a) plot
$a(z)$ (red-solid line) obtained from inverse engineering  with a final fiber length of $z_{f}=6$, as compared to the adiabatic reference (blue-dashed line) with $z_{f}=60$, satisfying the adiabatic condition (\ref{adiabaticcondition-beta}).
Then, the corresponding dispersion profile $\beta_2(z)$ obtained from Eq. (\ref{reduced-Ermakov-2}) is depicted in Fig. \ref{reac}(b), alongside the adiabatic reference. With these results,
the STA protocol remarkably compresses the soliton to a specific final width roughly $10$ times faster than the adiabatic approach. 
It is noteworthy that the values of $\beta_2 (0)=1$ and $\beta_2 (z_f)=0.3$ obtained by both approaches are consistent with each other. 
Finally, as demonstrated in the previous subsection, Fig. \ref{reac} displays the soliton self-compression dynamics under both STA-based (panel (c)) and adiabatic (panel (d)) dispersion profiles, 
where the superior performance of the STA-based system is evidently demonstrated.
 
\begin{figure}[t]
\centering
\hspace*{-.5cm}\begin{minipage}{0.4\linewidth}
  \includegraphics[width=\linewidth]{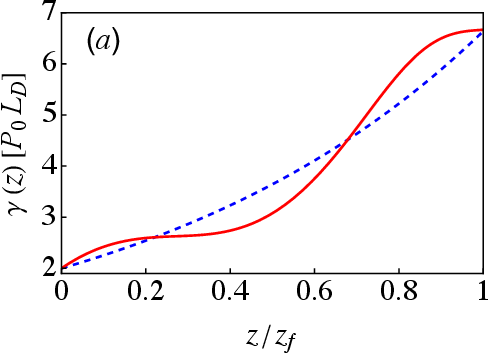}
\end{minipage}%
\hspace*{.5cm}\begin{minipage}{0.48\linewidth}
  \centering
  \includegraphics[width=\linewidth]{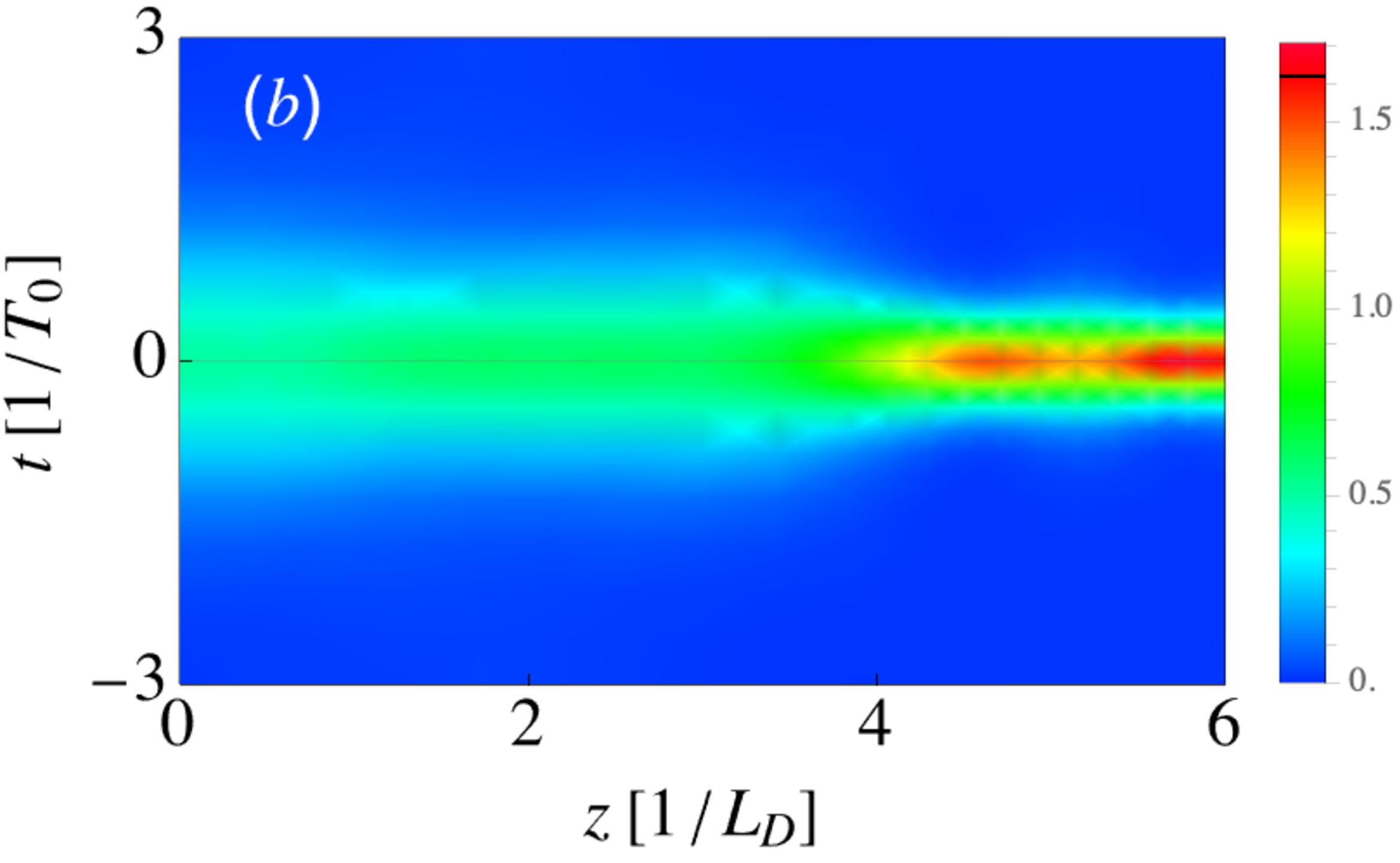}
\end{minipage}
\caption{(a) A comparison between the corresponding nonlinear function $\gamma(z)$ designed via inverse engineering (red-solid line) with $z_f=6$ and the adiabatic protocol $\gamma(z) = 2e^{\gamma_0 z}$ with constant $\gamma_0 =0.02$ (blue-dashed line) with $z_f=60$. (b) illustrates the spatio-temporal evolution of soliton wave packets in the STA protocol. Other parameters are: $\gamma(0)=2$, $\gamma(z_f)=6.6$, $\beta_2=1$, and $g=0$.}
\label{nonlinear}
\end{figure}


\subsection{Nonlinearity}
\label{nonlinearity}

In general, the interplay between dispersive and nonlinear effects yields the creation of solitons. As the pulse propagates along the fiber, the intensity-dependent correction to the refractive index induces self-phase modulation, an essential ingredient for soliton self-compression that leads to gentle symmetric spectral broadening. For a long nonlinear fiber, the strong cumulative interaction between the traveling pulse and the medium can bring additional (sometimes detrimental) effects such as four-wave mixing, self-steepening, stimulated Raman scattering, gain saturation, etc., which may significantly influence the efficiency of the self-compression process.

For the sake of completeness, let's consider the propagation of solitons within a nonlinear fiber characterized by constant GVD, while allowing the 
nonlinear parameters to vary without any gain or loss $g=0$. Thus, Eq. (\ref{ermakov}) can be reduced to 
\begin{equation}
\label{reduced-Ermakov-3}
 \ddot{a} = \frac{4 \beta^2_2}{\pi^2 a^3} - \frac{2 \beta_2 \gamma(z)}{\pi^2 a^2}.
\end{equation}
This equation allows for fast soliton self-compression, exceeding the adiabatic criteria, as discussed in previous references \cite{Koushik, Kong, Huangpra, Li2016}.

In this context, the nonlinearity $\gamma(z)$ serves the same role as $G(z)$,  the transformation of the distributed gain $g(z)$, as seen from Eqs. (\ref{reduced-Ermakov}) and (\ref{reduced-Ermakov-3}).
To avoid redundancy and facilitate a direct comparison with our prior results in Sec. \ref{gain}, 
we opt for an adiabatic protocol regarding the nonlinear parameter. Specifically, we employ an exponential profile for $\gamma(z) = 2e^{\gamma_0 z}$ with a constant $\gamma_0$. 
This choice establishes an adiabatic reference, denoted as $a_c(z) =2 \beta_2/\gamma(z)$. 
Similar to our method with distributed gain, we use a polynomial ansatz for the inverse engineering of the nonlinear function $\gamma(z)$, as depicted in Fig. \ref{nonlinear}(a). 
This is achieved by using a polynomial function Eq. (\ref{polynomial ansatz}) for $n=5$ with specific boundary conditions: $a(0) = a_c(0)$, $a(z_f) = a_c(z_f)$, and $\dot{a}(0) = \dot{a}(z_f) = \ddot{a}(0) = \ddot{a}(z_f) = 0$. 
During the propagation, the nonlinear function $\gamma(z)$ gradually increases at different distances, following distinct trajectories to accomplish soliton compression. 
The resulting evolution with the STA protocol is illustrated in Fig. \ref{nonlinear}(b). 
By selecting appropriate values for $\gamma(0)=2$ and $\gamma(z_f)=6.6$, we attain the results equivalent to the case of distributed gain, as $G(z)\gamma(z)$ remains the same in Eq. (\ref{ermakov}).
In a word, we present an alternative approach for achieving rapid self-compression by manipulating either the nonlinearity or the distributed gain. It's important to highlight that the latter method might require more intricate analysis due to the complex boundary conditions involved. Furthermore, there's potential for combining distributed gain and nonlinearity to further enhance soliton compression.


\section{Discussion}
\label{discussion}
\subsection{Stability}

In our exploration of the dynamical stability of soliton self-compression during the accelerated self-compression process, we introduce a measure of fidelity, 
denoted as $F=|\langle A^{ad}(z_{f},t)|A(z_{f},t)\rangle|^{2}$. Here, $|A(z_{f},t)\rangle$ represents the numerical output for the electric field evolved along the shortcut trajectory, while $|A^{ad}(z_{f},t)\rangle$ signifies the ideal electric field predicted by the adiabatic reference.

\begin{figure}[t]
\centering
\hspace*{-1cm}\includegraphics[width = 0.85\linewidth]{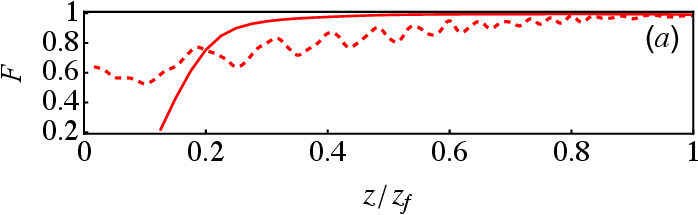}
\hspace*{-1cm}\includegraphics[width = .85\linewidth]{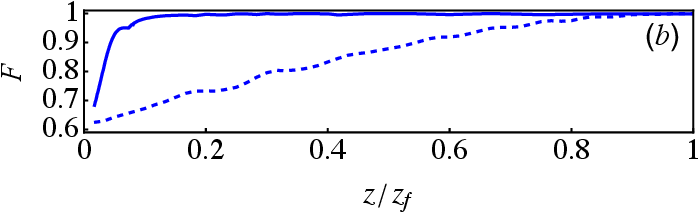}
{\hspace*{-1cm}\includegraphics[width = .85\linewidth]{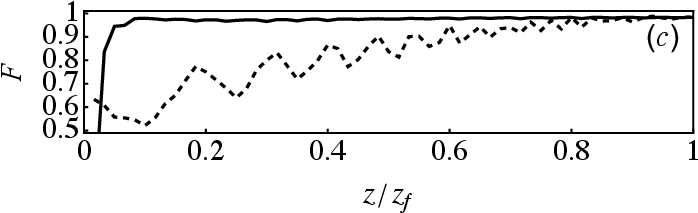}}
\caption{The fidelity versus propagation distance $z$ for STA (solid line) and adiabatic protocol (dashed line) soliton compression in three sections: (a) distributed gain, (b) decreasing GVD, and (c) variable nonlinearity strength. The corresponding parameters in different cases are the same as those in Fig.~\ref{evolutionone} to ~\ref{nonlinear}.}
\label{fidelity}
\end{figure}

As demonstrated in the preceding sections, the STA protocol excels in achieving precise soliton compression over shorter distances compared to adiabatic optimization strategies under three different controllable parameter designs. The obtained fidelity in the STA protocol is remarkably high, nearly perfect with values reaching up to 0.99, as indicated by the solid lines in Fig. \ref{fidelity}. On the other hand, the adiabatic process, represented by dashed lines in Fig. \ref{fidelity}, requires a longer propagation distance to achieve soliton compression. Furthermore, the adiabatic-driven dynamics exhibit strong oscillations in Fig. \ref{fidelity}(a), (c). These oscillations are associated with the specific forms of the parameter functions utilized in the adiabatic reference Eq. (\ref{ad}). Such oscillations might be mitigated by choosing alternative forms for the control functions $g(z)$ and $\gamma(z)$.

Additionally, implementation of the proposed protocol for compressing soliton requires careful consideration of the relationship between the different control parameters. For instance, to achieve soliton compression through inverse-engineered dispersion, nonlinearity and gain, one could utilize a fiber-ring-laser-like structure by dividing the fiber into three parts and using a mode-locked erbium fiber oscillator as the source of the bandwidth-limited solitons \cite{Peng}. To this end, one can use a short erbium-doped fiber to produce the desired gain profile (as shown in Fig. \ref{evolutionone}(b)) \cite{fiber_gain} whereas the nonlinear properties can be controlled using tapered photonic crystal fibers \cite{Tapers}. Subsequently, the pulses can be diverted into a DDF-based pulse compressor to obtain the required compression using a longitudinally-varying inverse-engineered dispersion profile.


\subsection{Third-order dispersion}

The above sections have analyzed soliton self-compression in fibers whose dispersion profile only comprises the GVD term. Although the contribution of this term dominates in most cases of practical interest, it is sometimes necessary to consider the TOD, $\beta_3(z)$ for soliton self-compression \cite{todGPA}, effects start to become significant when the condition $T_0|\beta_2/\beta_3| \le 1$ is satisfied \cite{sch}. For instance, considering a 10 ps pulse, this condition implies that $|\beta_2| = 10^{-2}$ $\text{ps}^2/\text{km}$ when $|\beta_3| = 0.1$ $\text{ps}^3/\text{km}$. To illustrate its effect, we include $\beta_3$ for the case of inverse-engineered gain to achieve soliton compression, as discussed in Sec. \ref{gain}. 
The modified NLSE now reads:
\begin{equation}
i\frac{\partial A}{\partial z}+\frac{1}{2}\beta_2\frac{\partial^{2}A}{\partial t^{2}}-\frac{i}{6}\beta_3\frac{\partial^{3} A}{\partial t^{3}}+\gamma G(z)|A|^{2}A=0 ,
\end{equation}
where $G$ can be presented by STA and adiabatic protocol, as described above. 

\begin{figure}[t]
\centering
\hspace*{-0.3cm}\includegraphics[width = 0.85\linewidth]{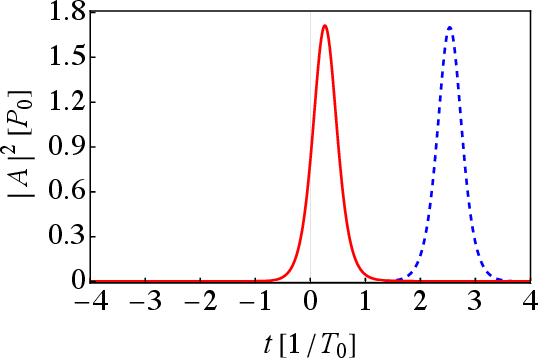}
\caption{Comparing the output intensity profile via STA protocol ($z_f =6$, red-solid line) and adiabatic protocol evolution ($z_f = 60$, blue-dashed line). Other parameters are: $\beta_3 = 0.006$, $\gamma=2$, $\beta_2=1$, and $g_0=0.01$ for adiabatic reference.}
\label{tod}
\end{figure}

Numerical modelling shows that, for our parameters, TOD mainly causes a temporal displacement of the centroid of the soliton along propagation (for shorter pulses in the 10 fs of femtoseconds range, the effect is more dramatic causing severe asymmetric distortions in the temporal profile \cite{sch}). This shift is much more pronounced in the system optimized by the adiabatic approach (blue-dashed line in Fig.~\ref{tod} as compared to the STA-based protocol red-solid line in Fig.~\ref{tod}). This is, again, mainly because of the longer propagation lengths required to achieve the same compression factors.

\subsection{Decompression}

We have demonstrated that a combination of the variational approximation and inverse engineering for distributed gain, decreasing dispersion, and variable nonlinear strength, enables fast and robust soliton self-compression. To complete the study, we have also analyzed the reverse process: soliton decompression or expansion. To do so, we have kept both dispersion and nonlinearity constant, and only modify the gain profile. One might naively expect that simply reversing the compression process would lead to efficient decompression. However, analysis of the pulse evolution revealed that the decompression process is not symmetric, meaning that it must be carefully considered on its own merit.

From Eq. (\ref{reduced-Ermakov}), we can easily get the expression of $G(z)$, $G(z)={1}/{a(z)}-{\ddot {a}(z)a^{2}(z)\pi^{2}}/{4}$. Again, one can engineer inversely $G(z)$ using a simple polynomial ansatz with fixed boundary conditions, as done before. Obviously, $G(z)$ must be greater than $0$, indicating:
\begin{equation}
\ddot{a}(z) < \frac{4}{\pi^{2}a^{3}(z)}.
\label{a_condition1}
\end{equation}
During the self-compression of the soliton, $\ddot{a}_{c}(z) < 0$ satisfies Eq. (\ref{a_condition1}) in the first half of the distance, which allows for the design of an ideal $g(z)$ within the short compression distance. However, as the soliton expands, the condition $\ddot{a}_{c}(z) > 0$ may not satisfy Eq. (\ref{a_condition1}). To address this issue, we use the mean-value theorem (MVT) \cite{pra_meanvalue} to set another useful bound for $z_{f}$. Since $a(z)$ is continuous in the interval $[0,z_{f}]$ and differentiable in $(0,z_{f})$, its maximum distance derivative must be limited by the MVT at point $z_f/2$. We can divide the interval $[0,z_{f}]$ in half, creating two symmetrical segments centred on this point, and then use the MVT again to obtain a lower bound for $z_{f}$.
\begin{equation}
z_{f}\geq\bigg(\pi^{2}a^{3}(z_{f})[a(z_{f})-a(0)]/2 \bigg)^{1/2}.
\label{zfbound}
\end{equation}
\begin{figure}
\centering
\includegraphics[width = 0.85\linewidth]{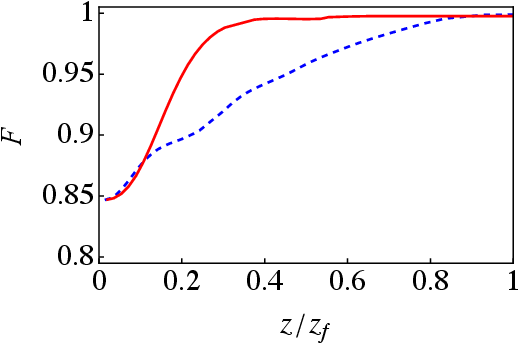}
  \caption{The fidelity versus propagation distance $z$ for STA ($z_f =25$, red-solid line) and adiabatic protocol ($z_f =70$, blue-dashed line) soliton compression. Other parameters are: $\beta_2=1$, $\gamma=2$, $a_c(0)=1$, $a_c(z_f)=2$, and $g_0=-0.005$ for the adiabatic reference.}
  \label{fid_loss}
\end{figure}

This provides a practical constraint on $z_{f}$ that ensures the validity of Eq. (\ref{a_condition1}) and facilitates the design of an appropriate $a(z)$ for the expansion phase of the soliton. It is important to note that MVT just provides a rough range of $z_{f}$ compared to the actual bound we found. This fact is shown in Fig.~\ref{fid_loss}, where it clearly shows that inverse engineered $a(z)$ can attain close to unit fidelity only when $z_{f}\geq 25$, which is obviously greater than the value we derived from Eq. (\ref{zfbound}), i.e., $7$. 

\section{Conclusion}
\label{conclusion}

In summary, we have investigated the control of the spatio-temporal evolution of bright solitary waves in active nonlinear Kerr media through the tailored variation of different physical parameters. We have derived analytical dynamical equations using the variational approximation and successfully designed smooth distributed gain, dispersion and nonlinearity profiles through inverse engineering via the STA protocol, 
achieving fast and high-fidelity self-compression. For completeness, we have also studied the expansion behaviour of solitons, which sheds light on the many possibilities of optimising soliton dynamics. Our results for this type of model is crucial for understanding and optimizing the behavior of optical pulses in complex media, which is essential for applications like high-speed communication and information processing. 
Future research could explore the potential benefits of combining STA with optimal control theory to achieve even more refined control over the soliton evolution. It must be emphasized that our results are not limited to any specific systems. We may consider a different and more generic design, e.g., for cubic-quintic (high-order) \cite{ZhangPRA} or nonlocal nonlinear \cite{Olebang} media \cite{Borispra,Agarwalpre,Sabari} as well. Instead, the methods developed here can be readily extended to the manipulation of other types of dissipative solitons \cite{Li,Peng} and even non-solitonic nonlinear waves such as similaritons \cite{similariton}.

\begin{acknowledgments}
 This work is supported by NSFC (12075145), EU FET Open Grant  EPIQUS (899368), the Basque Government through Grants No. IT1470-22 and IT1452-22 and ELKARTEK (KK-2021/00082 and KK-2021/00092), the China Scholarship Council (CSC) under Grant No.CSC N.202206890003, the project grants PID2021-126273NB-I00, PID2021-123131NA-I00 and TED2021-129959B-C21 funded by MCIN/AEI/10.13039/501100011033, by "ERDF a way of making Europe", and by the European Union NextGenerationEU/PRTR", and the IKUR Strategy of the Basque Government under the collaboration agreement between Ikerbasque Foundation and the University of the Basque Country. X.C. acknowledges ayudas para contratos Ram\'on y Cajal–2015-2020 (RYC-2017-22482). 
\end{acknowledgments}

\appendix

\section{Variational approximation approach}
\label{va}

Substituting the $ansatz$ in Eq. (\ref{ansatz}) into Eq. (\ref{sch_new}) yields the Lagrangian density:
\begin{equation}
\mathcal{L} = \frac{i}{2}\left(\frac{\partial A}{\partial{z}}A^{*}
-\frac{\partial A^{*}}{\partial{z}}A\right)
-\frac{\beta_2(z)}{2}\left|\frac{\partial A}{\partial{t}}\right|^{2}
+\frac{\gamma(z)}{2}G(z)| A|^{4}.
\label{lag}
\end{equation}
Through integration over $t$, $L=\int_{-\infty }^{+ \infty }\mathcal{L}dt$, we obtain the averaged Lagrangian
\begin{equation}
	\begin{aligned}
		L=&-\frac{A_0^2\beta_2(z)}{3a}+\frac{2aA_0^4G(z)\gamma(z)}{3 }-\frac{1}{3}\pi^{2}a^{3}A_0^2b^{2}\beta_2(z)
\\&-\frac{1}{6}\pi^{2}a^3A_0^2\frac{db}{dz}-2aA_0^2\frac{d\phi}{dz}.
	\end{aligned}
	\label{efflag}
\end{equation}
Using the Euler-Lagrange formulas, 
\begin{equation}
	\frac{\mathrm{d}}{\mathrm{d}z}\left(\frac{\partial L}{\partial \dot q}\right)-\frac{\partial L}{\partial q}=0,
\end{equation}
where $q$ represents one of the parameters $A(z), a(z), b(z)$ and $\phi(z)$, while $\dot{}$ indicates the derivative with respect to $z$. Taking variations of the averaged Lagrangian concerning the parameters, we obtain that
\begin{align} 
	\frac{\delta L}{\delta b}={}&\frac{1}{3}ab\beta_2(z)-\frac{1}{4}\frac{da}{dz}-\frac{1}{3}a^2A_0\frac{dA_0}{dz}=0,\label{1}\\
	\frac{\delta L}{\delta A}= {}&-\frac{4}{3}A_0^2G(z)\gamma(z)+\frac{1}{3}\pi^2a^2b^2\beta_2(z)\label{2}
	\\&+\frac{\beta_2(z)}{3a^2}+\frac{1}{6}\pi^2a^2\frac{db}{dz}+2\frac{d\phi}{dz}=0, \notag\\
	\frac{\delta L}{\delta a}={}&-\frac{2}{3}A_0^2G(z)g(z)+\pi^2a^2b^2\beta_2(z)\label{3}
	\\&-\frac{\beta_2(z)}{3a^2}+\frac{1}{2}\pi^2a^2\frac{db}{dz}+2\frac{d\phi}{dz}=0. \notag
\end{align}
A set of coupled differential equations can be derived from Eq. (\ref{1}) and Eq. (\ref{2}), produces the evolution equation for the soliton's parameters
\begin{eqnarray}
	\frac{da}{dz} &=& 2\beta_2(z)ab,\\
	\frac{db}{dz} &=& \frac{2\beta_2(z)}{\pi^{2}a^{4}}-2\beta_2(z)b^{2}-\frac{\gamma(z)G(z)}{\pi^{2}a^{3}}.
\end{eqnarray}
It should be noted that $\phi$ does not play any role in the variational dynamics and hence we put $\phi=0$ in the rest of the work. Above set of equations could be simplified to the following equation :
\begin{equation}
	\frac{d}{dz}\left[\frac{1}{\beta_2(z)}\frac{da}{dz}\right]=\frac{4\beta_2(z)}{\pi^{2}a^{3}}-\frac{2\gamma(z)G(z)}{\pi^{2}a^{2}}\equiv-\frac{\partial V}{\partial a}.
\end{equation}


\begin{thebibliography}{99}
	\bibitem{solitonrev} H. A. Haus and W. S. Wong, Solitons in optical communications, \href{https://link.aps.org/doi/10.1103/RevModPhys.68.423}{Rev. Mod. Phys. \textbf{68}, 423  (1996).}
	
	
	\bibitem{Agrawal} P. Y. S. Kivshar and G. P. Agrawal, \textit{
		Optical Solitons: From Fibers to Photonic Crystals}, San Diego: Academic Press 2012.
	
	\bibitem{sch} G. P. Agrawal, \textit{Nonlinear Fiber Optics}, New York: Academic Press 2013.

	\bibitem{shankapl} C. V. Shank, R. L. Fork, R. Yen, and R. H. Stolen, Compression of femtosecond optical pulses \href{https://doi.org/10.1063/1.93276}{Appl. Phys. Lett. \textbf{40}, 761 (1982).}
	
	\bibitem{ForkOL} R. L. Fork, C. H. Brito Cruz, P. C. Becker, and C. V. Shank,
	Compression of optical pulses to six femtoseconds by using cubic phase compensation, \href{https://doi.org/10.1364/OL.12.000483}{Opt. Lett. \textbf{12}, 483-485 (1987).}
	
	\bibitem{TomlinsonJOSA} W. J. Tomlinson and W. H. Knox, Limits of fiber-grating optical pulse compression, \href{https://doi.org/10.1364/JOSAB.4.001404}{J. Opt. Soc. Am. B \textbf{4}, 1404-1411 (1987).}
	
	\bibitem{Gordonprl} L. F. Mollenauer, R. H. Stolen, and J. P. Gordon,  Experimental Observation of Picosecond Pulse Narrowing and Solitons in Optical Fibers,
	\href{https://link.aps.org/doi/10.1103/PhysRevLett.45.1095}{Phys. Rev. Lett. \textbf{45}, 1095  (1980).}
	
	\bibitem{Tomlinsonol} L. F. Mollenauer, R. H. Stolen, J. P. Gordon, and W. J. Tomlinson, Extreme picosecond pulse narrowing by means of soliton effect in single-mode optical fibers, \href{https://doi.org/10.1364/OL.8.000289}{Opt. Lett. \textbf{8}, 289-291 (1983).}
	
	\bibitem{Ahmedieee} K.A. Ahmed, Kai Choong Chan, and Hai-Feng Liu, Femtosecond pulse generation from semiconductor lasers using the soliton-effect compression technique, \href{ 10.1109/2944.401246}{IEEE J. Select. Topics Quantum Electron. \textbf{1},  592 (1995).}
	
	\bibitem{FosterOE} M. A. Foster, A. L. Gaeta, Q. Cao, and R. Trebino, 
	Soliton-effect compression of supercontinuum to few-cycle durations in photonic nanowires, \href{https://doi.org/10.1364/OPEX.13.006848}{Opt. Express \textbf{13},  6848-6855 (2005).}
	
	
	\bibitem{Kuehl} H. H. Kuehl, Solitons on an axially nonuniform optical fiber, \href{doi:10.1364/JOSAB.5.000709}{J. Opt. Soc. Am. B \textbf{5}, 709 (1988).} 
	
	\bibitem{Smith1989} K. Smith and L. F. Mollenauer, Experimental observation of adiabatic compression and expansion of soliton pulses over long fiber paths, \href{https://doi.org/10.1364/OL.14.000751}{Opt. Lett \textbf{14}, 751-753 (1989).}
	
	\bibitem{ChernikovJOSA} S. V. Chernikov and P. V. Mamysh, Femtosecond soliton propagation in fibers with slowly decreasing dispersion, \href{https://doi.org/10.1364/JOSAB.8.001633}{J. Opt. Soc. Am. B \textbf{8}, 1633-1641 (1991). }
	
	\bibitem{ChernkovEL} S. V. Chernikov, Picosecond soliton pulse compressor based on dispersion decreasing fiber, \href{doi:10.1049/el:19921175}{Electron. Lett. \textbf{28}, 1842 (1992).} 
	
	\bibitem{Mamyshevprl} P. V. Mamyshev, P. G. J. Wigley, J. Wilson, G. I. Stegeman, V. A. Semeonov, E. M. Dianov, and S. I. Miroshnichenko, Adiabatic compression of Schrödinger solitons due to the combined perturbations of higher-order dispersion and delayed nonlinear response, \href{https://link.aps.org/doi/10.1103/PhysRevLett.71.73}{Phys. Rev. Lett. \textbf{71}, 73  (1993).}
	
	\bibitem{Chernikovol} S. V. Chernikov, E. M. Dianov, D. J. Richardson, and D. N. Payne, Soliton pulse compression in dispersion-decreasing fiber, \href{https://doi.org/10.1364/OL.18.000476}{ Opt. Lett. \textbf{18},  476-478 (1993).}
	
	\bibitem{Pelusi}  M. D. Pelusi, and H. F. Liu, Higher-order soliton pulse compression in dispersion-decreasing optical fibers, \href{10.1109/3.605567}{IEEE J. Quantum Electron \textbf{33}, 1430-1439 (1997).}

\bibitem{CPA55} D. H. Deng, L.  Zhan, Z. C. Gu, Y. Gu, and Y. X. Xia, 55-fs pulse generation without wave-breaking from an all-fiber Erbium-doped ring laser, \href{https://opg.optica.org/oe/fulltext.cfm?uri=oe-17-6-4284&id=177043}{Opt. Express \textbf{17}, 4284-4288 (2009).}

 
	\bibitem{DDFad} L. M. Ivanov, P.P. Branzalov, and L.I. Pavlov, Efficient compression of fundamental solitons in fibres with decreasing dispersion. \href{https://doi.org/10.1007/BF00619755}{Opt. Quant. Electron \textbf{24}, 565–573 (1992).} 
	
	\bibitem{Fatemi} F. K. Fatemi, Analysis of nonadiabatically compressed pulses from dispersion-decreasing fiber,  \href{doi:10.1364/OL.27.001637}{Opt. Lett. \textbf{27} , 1637 (2002).} 
	
	\bibitem{Travers} J. C. Travers, J. M. Stone, A. B. Rulkov, B. A. Cumberland, A. K. George, S. V. Popov, J. C. Knight, and J. R. Taylor, Optical pulse compression in dispersion decreasing photonic crystal fiber, \href{https://doi.org/10.1364/OE.15.013203}{Opt.  Express  \textbf{15},  13203-13211 (2007).}
	
	\bibitem{Lagsgaard} J. Lægsgaard and P. J. Roberts, Theory of adiabatic pressure-gradient soliton compression in hollow-core photonic bandgap fibers, \href{https://doi.org/10.1364/OL.34.003710}{Opt. Lett. \textbf{34}, 3710-3712 (2009).}
	
	\bibitem{Turitsynprl} A. Bednyakova, and S. K. Turitsyn, Adiabatic Soliton Laser, \href{https://link.aps.org/doi/10.1103/PhysRevLett.114.113901}{Phys. Rev. Lett. \textbf{114}, 113901 (2015).}
	
	\bibitem{Turitsyn} S. K. Turitsyn, 
	On the theory of adiabatic field dynamics in the Kerr medium with distributed gain and dispersion, \href{https://doi.org/10.1364/OL.44.001448}{Opt. Lett. \textbf{44},  1448-1451 (2019).}
	
	
	\bibitem{josb1994} D. Anderson, M. Lisak, B. Malomed, and M. Quirogateixeiro,
	Tunneling of an optical soliton through a fiber junction, \href{https://doi.org/10.1364/JOSAB.11.002380}{J. Opt. Soc. Am. B \textbf{11}, 2380-2384 (1994).}
		
	\bibitem{Koushik}  K. Paul, and A. K. Sarma,  Nonlinear compression of temporal solitons in an optical waveguide via inverse engineering, \href{https://iopscience.iop.org/article/10.1209/0295-5075/121/64001}{EPL \textbf{121}, 64001 (2018).}
	
	\bibitem{Kong}  Q. Kong, H. Ying, and X. Chen,  Shortcuts to Adiabaticity for Optical Beam Propagation in Nonlinear Gradient Refractive-Index Media, 
	\href{ https://doi.org/10.3390/e22060673}{Entropy \textbf{22}, 673 (2020).}

    \bibitem{fiber_non} A. Blanco-Redondo, C. Husko, D. Eades,  Y. Zhang, J. Li, T. F.  Krauss, and B.J. Eggleton, Observation of soliton compression in silicon photonic crystals, \href{https://doi.org/10.1038/ncomms4160}{Nat. Commun. \textbf{5}, 3160 (2014).}
 
	
	
	
	\bibitem{STArev}  D. Gu\'ery-Odelin, A. Ruschhaupt, A. Kiely, E. Torrontegui, S. Mart\'inez-Garaot, and J. G. Muga, Shortcuts to adiabaticity: Concepts, methods, and applications, \href{https://journals.aps.org/rmp/abstract/10.1103/RevModPhys.91.045001}{Rev. Mod. Phys. \textbf{91}, 045001 (2019)}.
	
	
	\bibitem{Zoller} V. M. P\'erez-Garc\'ia, H. Michinel, J. I. Cirac, M. Lewenstein, and P. Zoller, Low Energy Excitations of a Bose-Einstein Condensate: A Time-Dependent Variational Analysis, \href{https://link.aps.org/doi/10.1103/PhysRevLett.77.5320}{Phys. Rev. Lett. \textbf{77}, 5320 (1996).}
	
	\bibitem{Li2016} J. Li, K. Sun, and X. Chen,  Shortcut to adiabatic control of soliton matter waves by tunable interaction, \href{https://doi.org/10.1038/srep38258}{Sci. Rep. \textbf{6}, 38258 (2016).}
	
	\bibitem{Huangchaos} T. Y. Huang, B. A. Malomed, and X. Chen, Shortcuts to adiabaticity for an interacting Bose–Einstein condensate via exact solutions of the generalized Ermakov equation  \href{https://doi.org/10.1063/5.0004309
	}{Chaos: Interdiscip. J. Nonlinear Sci.  \textbf{30}: 053131 (2020).}
	
	\bibitem{Huangpra} T. Y. Huang, J. Zhang, J. Li, and X. Chen, Time-optimal variational control of a bright matter-wave soliton, \href{https://doi.org/10.1103/PhysRevA.102.053313}{Phys. Rev. A \textbf{102}: 053313 (2020).}
	
	\bibitem{Borisbook} Boris A. Malomed, Variational methods in nonlinear fiber optics and related fields, \href{https://www.sciencedirect.com/science/article/pii/S0079663802800269}{ \textit{Progress in Optics}, Vol. 43, Chapter 2, 71-193, Edited by E. Wolf,  Elsevier (2022).}
	
	\bibitem{Anderson} D. Anderson, Variational approach to nonlinear pulse propagation in optical fibers,  \href{{https://link.aps.org/doi/10.1103/PhysRevA.27.3135}}{Phys. Rev. A \textbf{27}, 3135 (1983).}
	
	\bibitem{Kruglovprl} V. I. Kruglov, A. C. Peacock, and J. D. Harvey, Exact self-similar solutions of the generalized nonlinear Schr\"odinger equation with distributed coefficients, \href{https://link.aps.org/doi/10.1103/PhysRevLett.90.113902}{Phys. Rev. Lett.  \textbf{90}, 113902 (2003).} 
	
\bibitem{ZhangPRA} J.-F. Zhang, Q. Tian, Y.-Y. Wang, C.-Q. Dai, and L. Wu, Self-similar optical pulses in competing cubic-quintic nonlinear media with distributed coefficients, \href{https://link.aps.org/doi/10.1103/PhysRevA.81.023832}{Phys. Rev. A \textbf{81}, 023832 (2012)}.

\bibitem{Olebang} W. Kr\'{o}likowski and O. Bang, Solitons in nonlocal nonlinear media: Exact solutions, \href{https://link.aps.org/doi/10.1103/PhysRevE.63.016610}{Phys. Rev. E \textbf{63}, 016610 (2000).}

\bibitem{Borispra} F. K. Abdullaev, J. G. Caputo, R. A. Kraenkel, and B. A. Malomed,
Controlling collapse in Bose-Einstein condensates by temporal modulation of the scattering length, \href{https://link.aps.org/doi/10.1103/PhysRevA.67.013605}{
	Phys. Rev. A \textbf{67}, 013605 (2003).}

\bibitem{Agarwalpre} R. Atre, P. K. Panigrahi, and G. S. Agarwal, Class of solitary wave solutions of the one-dimensional Gross-Pitaevskii equation, \href{https://link.aps.org/doi/10.1103/PhysRevE.73.056611}{Phys. Rev. E \textbf{73}, 056611 (2006).}

\bibitem{Sabari} S. Sabari and B. Dey,  Stabilization of trapless dipolar Bose-Einstein condensates by temporal modulation of the contact interaction, \href{https://link.aps.org/doi/10.1103/PhysRevE.98.042203}{Phys. Rev. E \textbf{98}, 042203 (2008).}

\bibitem{Tapers} E. C. M\"agi, P. Steinvurzel, and B. J. Eggleton, Tapered photonic crystal fibers \href{https://doi.org/10.1364/OPEX.12.000776}{Opt. Express \textbf{12}, 776-784 (2004).}

\bibitem{variablemass} P. G. L. Leach, Harmonic oscillator with variable mass \href{https://dx.doi.org/10.1088/0305-4470/16/14/019}{J. Phys. A \textbf{16}, 3261 (1983).}


\bibitem{nonlinear2003} V. I. Kruglov, A. C. Peacock, and J. D. Harvey, Exact self-similar solutions of the generalized nonlinear Schrödinger equation with distributed coefficients, \href{https://journals.aps.org/prl/abstract/10.1103/PhysRevLett.90.113902}{Phys. Rev. Lett. \textbf{90}, 113902 (2003).}

\bibitem{josb1996} M. L. Quiroga-Teixeiro, D. Anderson, P. A. Andrekson, A. Berntson, and M. Lisak, Efficient soliton compression by fast adiabatic amplification, \href{https://doi.org/10.1364/JOSAB.13.000687}{J. Opt. Soc. Am. B \textbf{13}, 687-692 (1996).}

\bibitem{DDLfiber} K. Tajima, Compensation of soliton broadening in nonlinear optical fibers with loss, \href{https://opg.optica.org/ol/abstract.cfm?uri=ol-12-1-54} {Opt. Lett. \textbf{12}, 54-56 (1987).}


\bibitem{DDFexp} J. S. Andrew, W. B. Robert, and F. E. Alan, Dramatically improved transmission of ultrashort solitons through 40 km of dispersion-decreasing fiber, \href{https://opg.optica.org/ol/abstract.cfm?URI=ol-20-17-1770}{Opt. Lett. \textbf{20}, 1770-1772 (1995).}

\bibitem{Peng} J. S. Peng, S. Boscolo, Z. H, Zhao, and H. P. Zeng, Breathing dissipative solitons in mode-locked fiber lasers, \href{https://www.science.org/doi/10.1126/sciadv.aax1110}{Sci. Adv \textbf{5}, eaax1110 (2019).}
\bibitem{fiber_gain} K. Tamura, E. P. Ippen, H. A. Haus, and L. E. Nelson, 77-fs pulse generation from a stretched-pulse mode-locked all-fiber ring laser,\href{https://opg.optica.org/ol/abstract.cfm?uri=ol-18-13-1080} {Opt. Lett. \textbf{18}, 1080-1082 (1993).}

\bibitem{todGPA} T. I. Lakoba, and G. P. Agrawal, Effects of third-order dispersion on dispersion-managed solitons, \href{https://opg.optica.org/josab/abstract.cfm?uri=josab-16-9-1332}{J. Opt. Soc. Am. B \textbf{16}, 1332-1343 (1999).}

\bibitem{pra_meanvalue} E. Torrontegui, S. Ib\'a\~nez, X. Chen, A. Ruschhaupt, D. Gu\'ery-Odelin, and J. G. Muga, Fast atomic transport without vibrational heating, \href{https://journals.aps.org/pra/abstract/10.1103/PhysRevA.83.013415}{Phys. Rev. A \textbf{83}, 013415 (2011)}

\bibitem{Li} D. Li, L. Li, J. Y.  Zhou, D. Y. Tang, and D. Y. Shen, Characterization and compression of dissipative-soliton-resonance pulses in fiber lasers,  \href{https://doi.org/10.1038/srep23631}{Sci. Rep. \textbf{6}, 23631 (2016).}

\bibitem{similariton} S. A. Ponomarenko and G. P. Agrawal, Optical similaritons in nonlinear waveguides,  \href{https://doi.org/10.1364/OL.32.001659}{Opt. Lett. \textbf{32}, 1659-1661 (2007).}

\end{thebibliography}
\end{document}